\documentclass[sigconf]{acmart}
\usepackage{graphicx}

\usepackage{booktabs} 
\usepackage[linesnumbered,boxed,ruled]{algorithm2e}
\usepackage{graphicx, subfigure}
\usepackage{float}
\usepackage{enumitem}
\usepackage{comment}%
\usepackage{amsmath,amssymb}

\usepackage{array}
\newcommand{\PreserveBackslash}[1]{\let\temp=\\#1\let\\=\temp}
\newcolumntype{C}[1]{>{\PreserveBackslash\centering}p{#1}}
\newcolumntype{R}[1]{>{\PreserveBackslash\raggedleft}p{#1}}
\newcolumntype{L}[1]{>{\PreserveBackslash\raggedright}p{#1}}

\usepackage{balance}

\def\BibTeX{{\rm B\kern-.05em{\sc i\kern-.025em b}\kern-.08emT\kern-.1667em\lower.7ex\hbox{E}\kern-.125emX}}
    
\copyrightyear{2019} 
\acmYear{2019} 
\acmConference[CIKM '19]{The 28th ACM International Conference on Information and Knowledge Management}{November 3--7, 2019}{Beijing, China}
\acmBooktitle{The 28th ACM International Conference on Information and Knowledge Management (CIKM '19), November 3--7, 2019, Beijing, China}
\acmPrice{15.00}
\acmDOI{10.1145/3357384.3357835}
\acmISBN{978-1-4503-6976-3/19/11}

\begin{document}

\fancyhead{}

\title{Graph Representation Learning for Merchant Incentive Optimization in Mobile Payment Marketing}

\author{Ziqi Liu$^\star$, Dong Wang$^\star$, Qianyu Yu, Zhiqiang Zhang, Yue Shen, \\
Jian Ma, Wenliang Zhong, Jinjie Gu, Jun Zhou, Shuang Yang, Yuan Qi
}\thanks{$^\star$Equal contribution.}
\affiliation{%
  \institution{Ant Financial Services Group}
  \city{Hangzhou}
  \country{China}
}
\email{{ziqiliu,yishan.wd,qianyu.yqy,lingyao.zzq,zhanying,mj.mj,yice.zwl,jinjie.gujj,jun.zhoujun,shuang.yang,yuan.qi}@antfin.com}
%
\renewcommand{\shortauthors}{Liu and Wang, et al.}

%
\begin{abstract}
Mobile payment such as Alipay has been widely used 
in our daily lives. To further promote the mobile payment activities, 
it is important to run marketing campaigns under a limited budget by 
providing incentives such as coupons, commissions to merchants.
As a result, incentive optimization is the key to maximizing the 
commercial objective of the marketing campaign.
With the analyses of online experiments, we 
found that the transaction network can subtly describe 
the similarity of merchants' responses to different 
incentives, which is of great use in the incentive optimization problem.
In this paper, we present a graph representation learning 
method atop of transaction networks for merchant
incentive optimization in mobile payment marketing. With limited samples 
collected from online experiments, our end-to-end method first learns 
merchant representations based on an attributed transaction networks, 
then effectively models the correlations between the commercial 
objectives each merchant may achieve and the incentives under varying
treatments. Thus we are able to model the sensitivity to incentive
for each merchant, and spend the most budgets on those merchants that 
show strong sensitivities in the marketing campaign.
Extensive offline and online experimental results at Alipay demonstrate 
the effectiveness of our proposed approach.
\end{abstract}

\ccsdesc[500]{Computing methodologies~Neural networks}
\ccsdesc[300]{Applied computing~Marketing}
%
\keywords{incentive optimization; mobile payment marketing; graph representation learning}

%

%
\maketitle

\section{Introduction}
In recent years, \emph{mobile payment services} (e.g., Alipay 
Pay~\footnote{\url{https://en.wikipedia.org/wiki/Alipay}} operated by 
Ant Financial, or Apple Pay~\footnote{\url{https://en.wikipedia.org/wiki/Apple\_Pay}} 
operated by Apple, etc.) have been playing much more important roles in users' daily lives. 
One major commercial goal of these mobile payment operators is to attract 
more merchants and customers to engage in their mobile payment services, 
instead of using traditional payment services. 

To encourage certain mobile payment activities, the operators of the payment services
would need to launch effective marketing campaigns. One way of promoting the payment 
activities is to offer the merchants with some kinds of incentives, e.g. commissions, 
coupons, value-added services and so on, when the merchants can attract 
customers to pay through the specific payment services.

\begin{figure}
\centering
\includegraphics[width=0.48\textwidth]{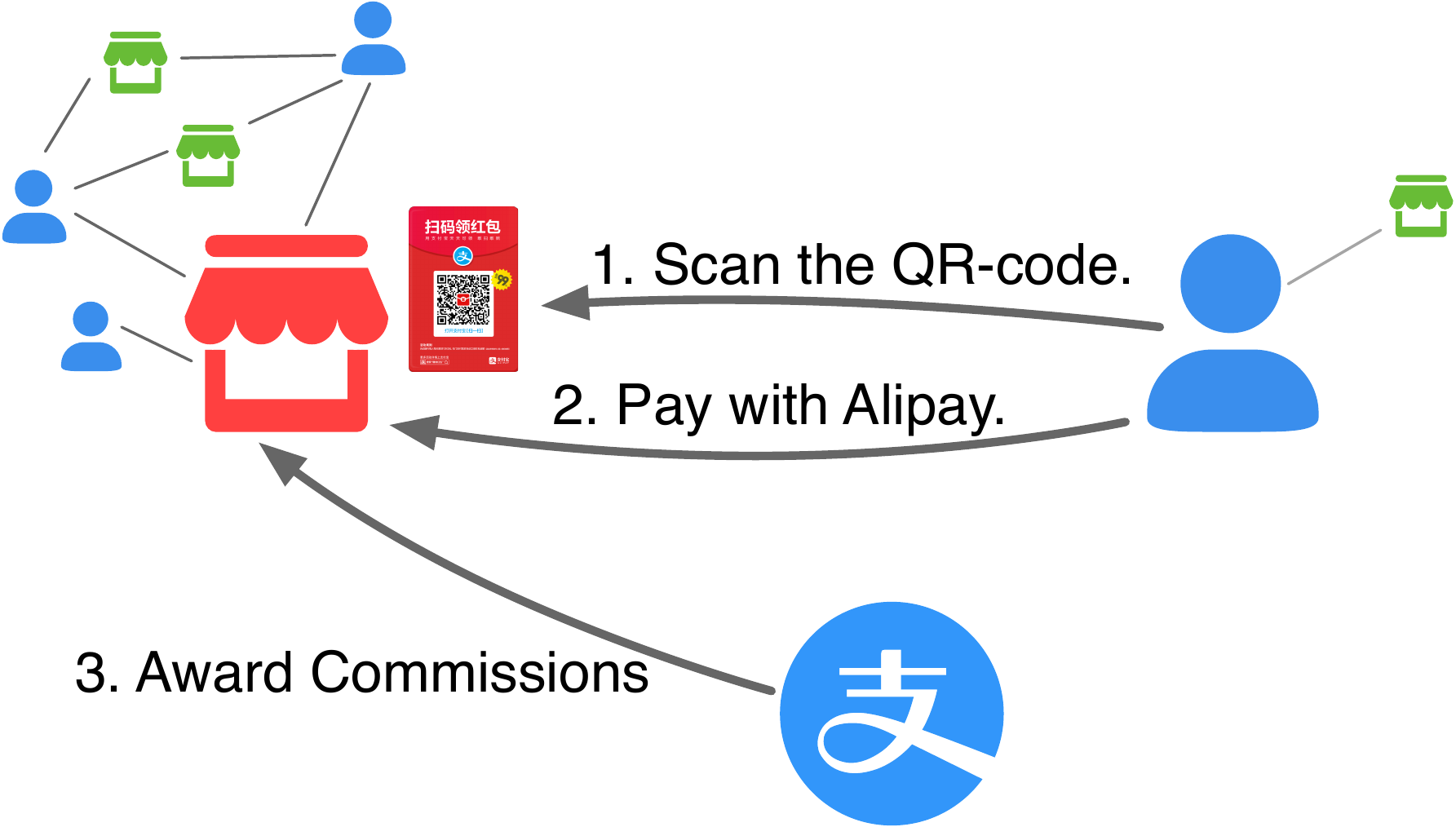}
\caption{An illustration of the Alipay marketing campaign. The customer
first scans the incentive QR-Code placed by merchant $i$, then redeem the incentive through 
a proper payment with any eligible merchant through Alipay. Finally Alipay will award operation 
commissions to merchant $i$. The more such payments a merchant can attract, the more it
will be awarded.} \label{fig:qrcode}
\end{figure}

Taking the above \emph{mobile payment marketing} as a concrete example at Alipay, each 
merchant of Alipay is assigned with a unique \emph{incentive QR Code}\footnote{Check
\url{https://intl.alipay.com/doc/redpacket/ctmz6b} for more details. The incentive
in this campaign is in the form of operation commissions. The merchant 
can earn an operation commission from each customer's redemption of the Alipay Red 
Packet QR Code (a specific form of incentive QR Code).} 
and is encouraged to ask their customers to scan,
so as to share the incentives to the customers.
The customers can redeem the shared incentives, similar to any other credit card points,
only by making a proper payment through
Alipay. At the same time, each successful redemption made by the customer
can award an equivalent amount of incentives as commissions to 
the merchant who shares the incentive.
With the incentives propagated and redeemed under such a 
mechanism (i.e. incentives first be shared from 
merchants to customers, then be redeemed through a proper payment), 
the more proper payments customers can accomplish, the more commissions merchants can gain.
We illustrate the marketing campaign in Figure~\ref{fig:qrcode}.


The amount of commissions assigned to merchants are simply rule-based in
traditional marketing strategies~\cite{clow2004integrated}.
However, in practice different merchants could response differently in a 
marketing campaign. For instance, for merchants with limited enthusiasms or abilities,
the number of payments they can attract may remain unchanged
no matter what amount of commissions they get, while others may show strong \emph{positive}
correlations between the commissions they acquired and the number of 
payments they can attract. That is, the 
merchants' \emph{sensitivities to the incentives} could vary a lot.
On the other hand, the budget of a marketing campaign is always limited. 
Therefore, in order to maximize the commercial objective of a payment marketing 
campaign under a limited budget, \emph{incentive optimization} is 
the key to success. To summarize, the operators should be able to target the 
merchants who show strong positive correlations between incentives 
(e.g. commissions) and objectives (e.g. the number of payments), and spend more budgets 
on those sensitive merchants.

However, modeling the correlations between incentives and objectives for
each merchant is non-trivial. The effect of the treatment\footnote{We define
one treatment in our case as a choice of a certain amount of incentive 
selected from a set of candidates. For example, award ``1 Chinese Yuan 
for every proper payment the merchant has attracted'' is a treatment in case
we have a set of candidates, say $\{1, 2, 5, 10, 20\}$ in Chinese Yuan.} on each merchant 
should be observed in a long
term because in practice each merchant could only be aware of the treatment based on 
an accumulation of incentives in a long enough period. Due to the limited budgets,
what we can do is to conduct ``treatment-limited'' online experiments by assigning a 
fixed treatment (say 1 Chinese Yuan) randomly sampled from a set of limited treatments 
$\{1, 2, 5, 10, 20\}$) 
to each merchant, and to observe the objectives (say the number of
payments) the merchant can attract after then. That is, it is 
infeasible to collect the whole objective-incentive curve empirically for 
any specific merchant (i.e. all the objectives one merchant can achieve
under all the treatments $\{1, 2, 5, 10, 20\}$). As such, we summarize the 
following two challenges. First, we require a 
representation learning method that can help embed merchants with 
similar ``sensitivities to the incentives''
to the similar representations in the vector space,
thus can help collect statistical significant data for estimating the objective-incentive
curve for a group of similar merchants. Second, 
the objective-incentive curve of each merchant we need to estimate could be 
arbitrarily complex in well-defined function spaces~\cite{hofmann2008kernel,tibshirani2014adaptive}. 
The variance of the estimation could be large if we do not place any 
reasonable priors on our model.

In this paper, we present a graph representation
learning approach to incentive optimization of merchants in the payment 
marketing scenario.
First, based on the intuition that the customers who interacted with are
strong signals to indicate whether a merchant has high potentials to 
promote the desired payments or not, we analyze the sensitivities of 
merchants located in different regions in section~\ref{sec:gg}. 
This motivates us to characterize
each merchant by considering the distribution of customers who made 
payments with the merchant, thus resort to
a variant of classic graph representation learning approach.
Second, we analyze the objective-incentive curves using random samples from 
real data, and show that the curve should lie in linear, monotonic function spaces.  
That is, if we let the number of transactions be the commercial objective, the number of 
transactions should increase linearly with the incentives.
Extensive experiments on real data show that our approach is effective.
Finally, we formalize the final decision making process under 
certain budget (assign the optimal amount of incentive to each merchant) as 
a linear programming problem, and show our online experimental results at Alipay.

\section{Background}
In this section, we briefly discuss related literatures to our problems.
Basically, we would first talk about related problems such as price 
optimization, and then review the literatures related to graph representation
learning approaches that forms the basis of our approach.

\subsection{Price Optimization}
Basically, price optimization uses data analysis techniques to address two problems.
(1) Understanding how customers will react to different pricing strategies for 
products and services. (2) Finding the best prices for a given company, 
considering its commercial goals.
Price optimization techniques can help retailers evaluate the potential 
impact of sales promotions or estimate the right price for each product 
if they want to sell it in a certain period of time.

Basically, the approaches to price optimization consist of two stages.
(1) Utilizing machine learning techniques to estimate sales quantity
of products~\cite{ito2017optimization}, sale probabilities from
partially observable market data~\cite{schlosser2018dynamic},
or historical lost sales and predict future demand of new 
products~\cite{ferreira2015analytics}. (2) With those estimates,
one can formalize optimization problems for the commercial goals
to get the optimal strategies.

Similar to price optimization, mobile payment services play the similar
roles as retailers, and aim to promote their products, i.e. specific
payments in our cases. Most of existing works exploit linear models
to estimate the sales~\cite{gallego2014multiproduct,ito2017optimization}.
In our setting we are required to estimate the sensitivities to
different treatments for a huge number of merchants with limited samples, 
which have few related works to our best of knowledge. 
We contribute a brand new model for estimating the sensitivities
of each merchant (product).

\subsection{Graph Representation Learning}
In this section, we review the literatures related to graph representation 
learning~\cite{hamilton2017representation}, especially graph neural networks
approaches that aim to encode the patterns of nodes' subgraphs as latent features.

Assuming an undirected graph $\mathcal{G}=(\mathcal{V}, \mathcal{E})$
with $N$ nodes $i \in \mathcal{V}$, $|\mathcal{E}|$ edges $(i, j) \in \mathcal{E}$,
the sparse adjacency matrix $A \in \mathbb{R}^{N\times N}$,
a matrix of node features $X \in \mathbb{R}^{N \times P}$, 
and graph Laplacian operator $L = I - D^{-1/2}A D^{-1/2}$~\cite{chung1997spectral}.
The approaches aim to learn to aggregate a subgraph of neighbors
associated with each target node.
For instance, Hamilton et al.~\cite{hamilton2017inductive}
proposed GraphSAGE, that defines a set of aggregator functions
in the following framework:
\begin{align}
  H^{(t+1)} = \sigma\bigg( \mathrm{CONCAT}\Big(\phi(A, H^{(t)}), H^{(t)}\Big) W^{(t)} \bigg), 
\end{align}
where $H^{(t)} \in \mathbb{R}^{N \times K}$ denotes the $t$-th hidden
layer with $H^{(0)}=X$, $W^{(t)}$ is the layer-specific
parameters, $\sigma$ denotes the activation function,
and $\phi(\cdot)$ is a pre-defined aggregator function over the graph,
that could be mean, max pooling operators.
By stacking the graph aggregation layers $T$ times, each node can
aggregate signals from its neighbors to $T$ hops.

To alleviate the limited representation capacity by doing aggregation 
only on ``pre-defined'' subgraphs~\cite{liu2018geniepath}, 
Veli{\v{c}}kovi{\'c} et al.~\cite{velivckovic2017graph} 
proposed attention based mechanisms to parameterize the aggregator functions.
Liu et al.~\cite{liu2018geniepath} propose an adaptive path layer
to further adaptively filter the receptive field of each node while
doing aggragation on the subgraphs.
Such methods have proven to achieve state-of-the-art results on several
public datasets. For users who are interested in the detailed literatures
please follow the comprehensive surveys~\cite{wu2019comprehensive}.

In our scenario, we exploit graph neural networks to embed each merchant 
by their customers who made payments in the transaction networks.
Such encoding can help describe the characteristics of customers each 
merchant makes transactions with, so as to evaluate the ability a merchant
may possess, and to describe the similarity of merchants’ responses 
to different incentives. Intuitively, merchants that have diverse customers
could possibly have much stronger abilities to attract more customers
compared with merchants that have limited customers. Two merchants
physically located in the same region should share similar customers,
thus they may have similar enthusiasms to share the incentives.
Such patterns can be well modeled based on the graph learning approaches
atop of the transaction network discussed later.

\section{Our Methodology}
In this section, we first introduce our online experiments to collect 
samples. After then, we present a variant of graph neural networks
for incentive optimization. Finally, we formalize the optimal strategy 
of treatments as a linear programming problem, based on the estimates
of merchants' personalized objective-incentive curve (or say sensitivities
to the marketing campaign), in a limited budget online setting.

\subsection{Online Experiments and Analyses}\label{sec:samples}
We aim to target those merchants who are able to boost the future commercial objectives
(e.g. the number of transactions using Alipay in our case)
as the operators stimulate greater incentives. That is, we need to discern merchants
with different sensitivities to the incentives. To obtain the sensitivities of each merchant
to the marketing campaign, we need to estimate the so called objective-incentive curve 
of each merchant.  The curve describes the mapping between incentives under 
different treatments and the commercial objectives that a merchant can achieve after 
receiving the treatment. Accurate estimations of the curves can help operators
to make decisions by choosing the best strategies.

\begin{figure}
\centering
\includegraphics[width=0.42\textwidth]{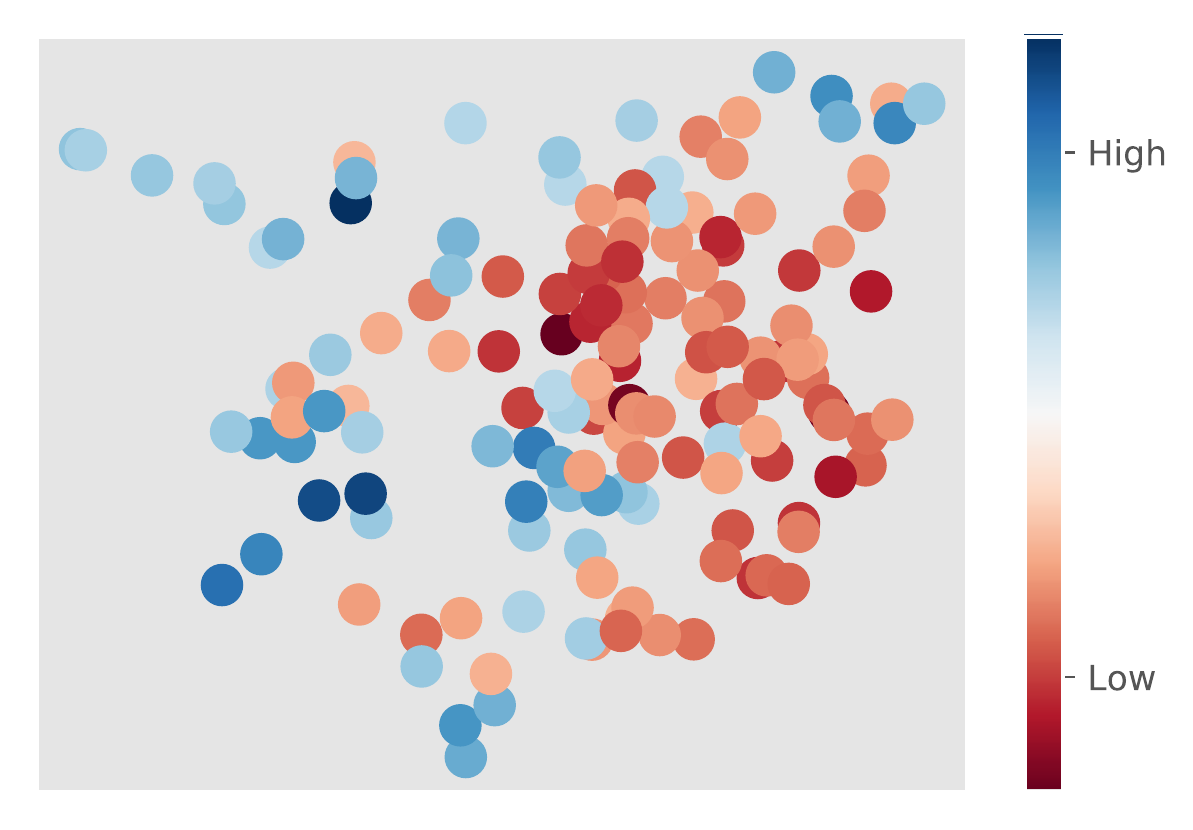}
\caption{The sensitivities of regions. The x-axis denotes the longitude, and y-axis denotes
the latitude. The colors marked with ``High'' indicates larger sensitivities compared with 
colors marked with ``Low''.} \label{fg:illu-geo}
\end{figure}

We choose to launch online experiments by randomly choosing a certain treatment for each
merchant in the online environment at Alipay, and observe the 
commercial objectives each merchant can achieve. The online experiment
involves millions of merchants and lasts for several days. The possible treatment 
is sampled from a set of incentives in Chinese Yuan. Hence, we can assume that 
the merchants who show similar sensitivities should be uniformly sampled
over all the treatments.

However, we can only observe one merchant's future objective under one treatment,
while we are aiming to estimate the whole objective-incentive curve of each merchant.
That means we need to represent merchants with similar capacities 
and enthusiasms similarly, and have to place
appropriate priors to constrain the functions estimating the curve.

\begin{figure*}
\centering
\includegraphics[width=0.9\textwidth]{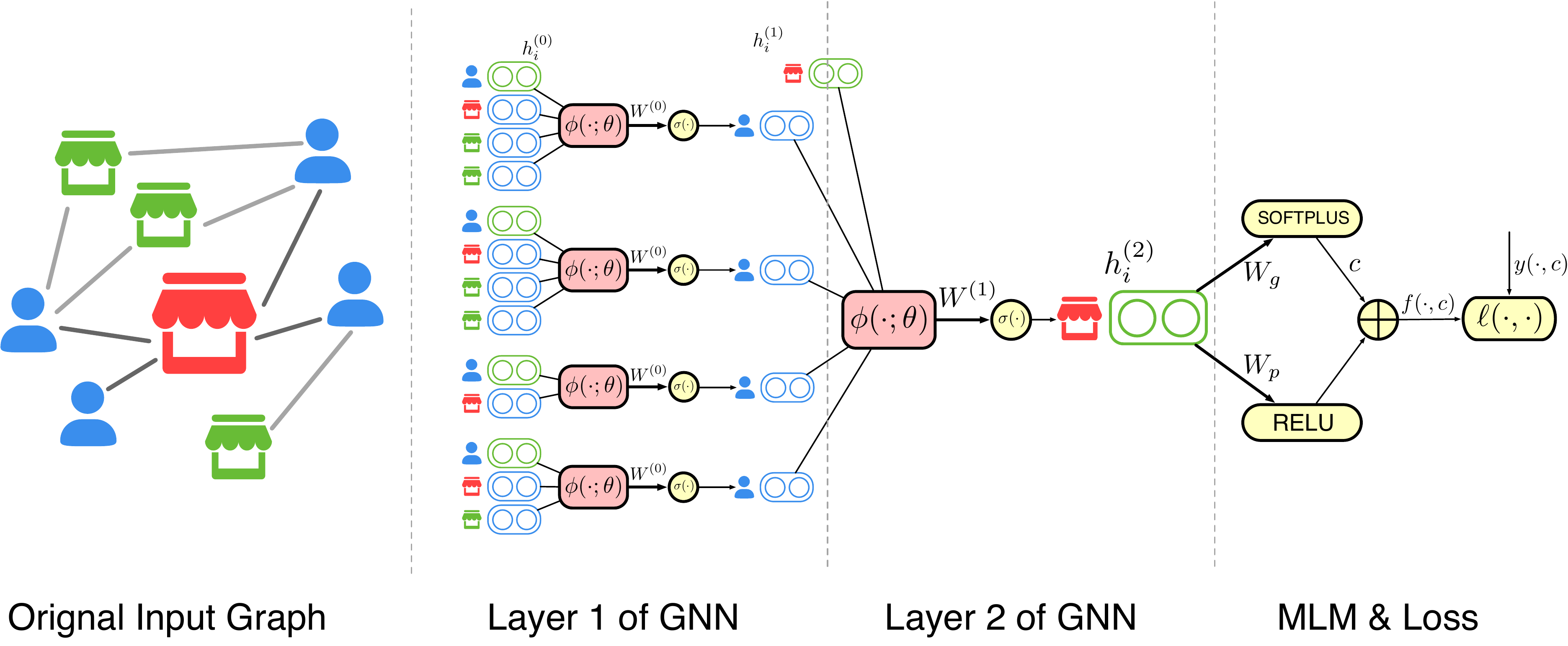}
\caption{The architecture of our graph neural networks with monotonic linear mapping (MLM)
in the output layer.} \label{fig:arch}
\end{figure*} 

\begin{figure}
\centering
\includegraphics[width=0.4\textwidth]{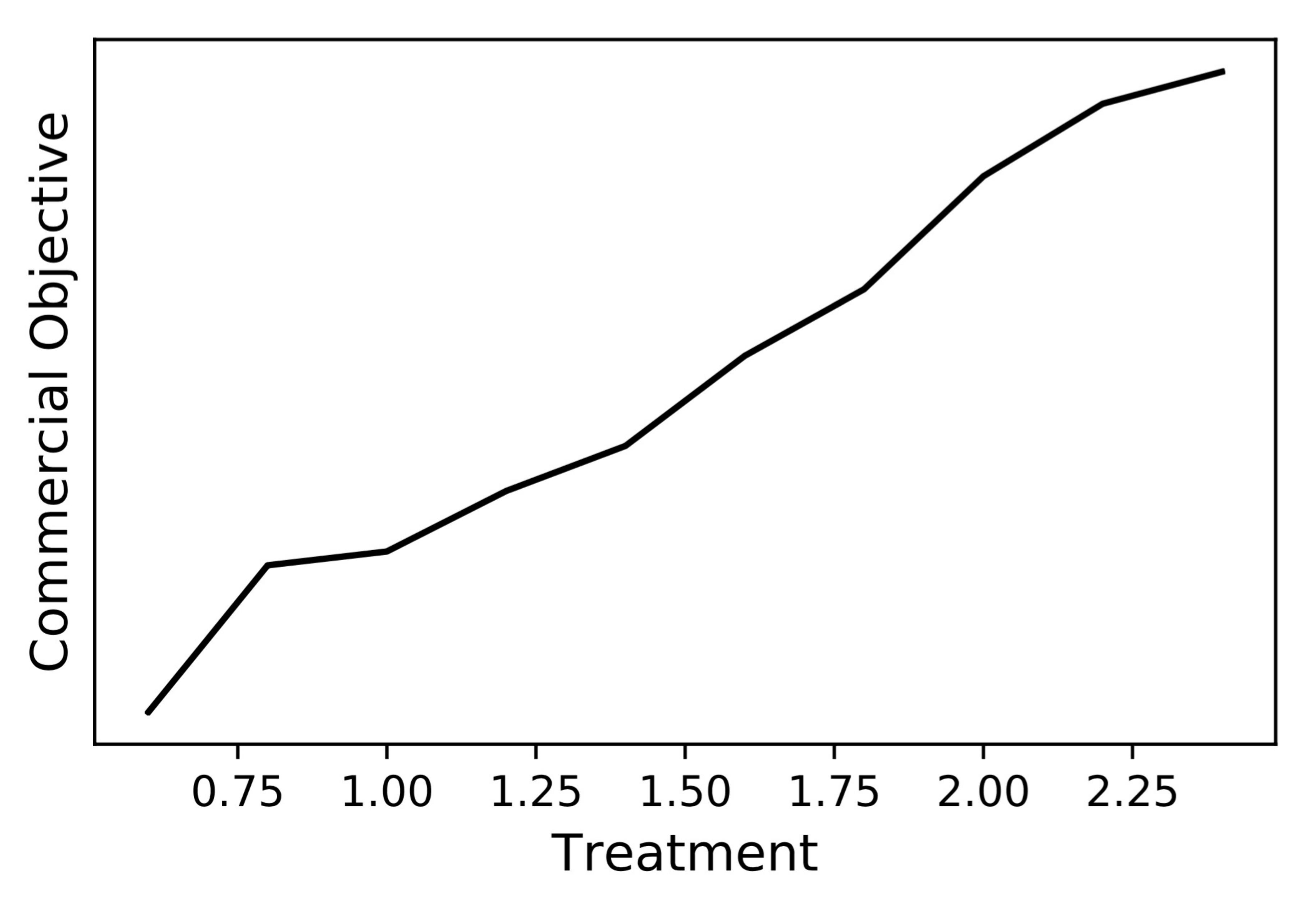}
\caption{Commercial objectives vary with treatments (each point 
denotes the average of \# payments (y-axis) over merchants under a given treatment (x-axis)
). 
} \label{fig:curve}
\end{figure}

First, the abilities of propagating incentives and attracting
proper payments can be naturally described through the distribution of customers 
with which the merchants are interacting. For example, merchants that have
transactions with diverse customers
are more competitive compared with merchants who have limited customers.
Merchants make transactions with the same group of customers could possibly be located
together, hence their enthusiasms of marketing the payment services
may be similar. 
To illustrate the relations between locations and sensitivities
of merchants, we group the merchants according to regions, and 
calculate the ``sensitivities of regions'' denoted by $AP(3.0, c)/AP(0.6, c)$,
where $AP(t, c)$ means the number of payments averaged over all merchants
from region $c$ with commission $t$. We plot the ``sensitivities of regions''
in figure~\ref{fg:illu-geo}. The figure shows that adjacent regions tend to 
share similar sensitivities. This led us to a representation learning 
approach based on graph neural networks atop of transaction networks.

Second, we show the expected objective-incentive curve using observations
from our online experiments in Figure~\ref{fig:curve}.
The figure shows that the expected mapping lies in a linear, monotonic function
space. In the next section, we will model such mapping using a linear transformation
given the representation of each merchant, where the derived ``gradient'' will be naturally 
used to characterize the sensitivity of each merchant.

\subsection{Models}
In this section, we present our models for incentive optimization based on 
the samples observed from our online experiments.
Estimating the objective-incentive curve for each merchant is non-trivial because
this requires a huge number of samples. We propose two methods to alleviate
the potential high variance in estimation. First, we resort to graph neural network based 
methods, and hopefully can embed merchants with similar 
characteristics to similar embeddings in the vector spaces, thus results into
statistical significant amount of observations with various treatment for 
a group of similar merchants. Second, we propose a monotonic linear transformation 
to map the graph embeddings learned for each merchant to the final 
objective-incentive curve. Our end-to-end model jointly
optimizes the graph neural networks and the monotonic linear transformation.

\subsubsection{Merchant Embedding based on Graph Neural Networks}\label{sec:gg}
In this part, we introduce how to embed merchants based on transaction graphs.

As we discussed above in section~\ref{sec:samples}, the merchants' distribution of customers could be 
represented from different dimensions, e.g. the location, the diversity of customers,
and so on. Therefore, we build a graph based on payments to connect
merchants and customers.

\begin{table*}
  \centering
  \caption{Experimental Dataset summary.}
  \label{tb:data}
  \begin{tabular}{ccccccc}
    \toprule
    Dataset & $\|V\|$ & $\|E\|$ & \# Node feature dim & \# Edge feature dim & \# Labeled merchants\\
    \midrule
    Dataset 1 & $90.08 \times 10^6$ & $161.7 \times 10^6$ & $4998$ & $86$ & $2.18 \times 10^6$ \\
    Dataset 2 & $97.18 \times 10^6$ & $172.1 \times 10^6$ & $4998$ & $86$ & $2.31 \times 10^6$ \\
  \bottomrule
\end{tabular}
\end{table*}

Assuming an undirected graph $\mathcal{G} = (\mathcal{V}, \mathcal{E})$ with $N$ nodes
$i \in \mathcal{V}$, $|\mathcal{E}|$ edges $(i,j) \in \mathcal{E}$. Each node
could be a merchant, a customer, or plays both roles. Node $i$ has a neighbor
$j$ (i.e. there exists an edge $(i,j)$) if there exists a proper payment where 
j acts as customer and i as merchant in the past several days. We assume the 
equivalent adjacent matrix $A \in 
\{0,1\}^{N \times N}$ of graph $\mathcal{G}$, where $A_{i,j}=1$ means 
there is an edge $(i,j)$ existed in graph $\mathcal{G}$.
Assuming $X \in \mathbb{R}^{N \times P}$ as the nodes' features
in $P$ dimensions. Assuming $Z \in \mathbb{R}^{|\mathcal{E}| \times D}$ denotes
the $D$-dimensional features, describing the payment related informations 
through the edges. We have the following graph neural network layers
to iterate $T$ times (thus propagate neighbors' signals in $T$ hops), 
and have the output encoding for node $i$ as $h_i^{T} \in \mathbb{R}^{K}$:
\begin{align}
  h_i^{(t+1)} = \sigma\bigg( {W^{(t)}}^\top \phi\bigg(\Big\{h_j^{(t)} | j 
\in \mathcal{N}(i) \cup i\Big\}; \theta \bigg)\bigg), 
\end{align}
where $h_i^{(t)} \in \mathbb{R}^{K}$ denotes node $i$'s itermediate embedding 
at the $t$-th hidden layer, $\sigma$ denotes the 
activation functions, and $\phi(\cdot)$ is the aggregator function parameterized 
by $\theta$ as defined in~\cite{liu2018geniepath}; 
$h_i^{(0)}={W_x}^\top X_i + 
\sum_{j \in \mathcal{N}(i)} {W_e}^\top Z_{(i,j)}$, $W^{(t)} 
\in \mathbb{R}^{K \times K}$, $W_x^{(t)} \in \mathbb{R}^{P \times K}$, and $W_e^{(t)} \in 
\mathbb{R}^{D \times K}$ are the parameters. Note that for each merchant our graph neural networks
not only capture the distribution over neighbors (i.e. distribution of merchants' customers), 
but also the distribution over edges (i.e. the distribution of payments between the merchant 
and its customers).

\subsubsection{Monotonic Linear Mapping}\label{sec:monotonic}
Given the embedding of each merchant based on graph neural networks convolved
on the transaction networks, we are able to model the objective-incentive
curves. Based on the analyses in section~\ref{sec:samples}, we have the following 
mapping:
\begin{align}\label{eq:mapping}
f(i, c) = c \cdot \underbrace{\mathrm{SOFTPLUS} 
(W_g^\top h_i^{(T)})}_{\text{gradient}} + \underbrace{\mathrm{RELU}(W_p^\top h_i^{(T)})}_{\text{intercept}}
\end{align}
where $f(i,c)$ denotes the estimation of the future objective of merchant $i$
under treatment $c$, and $W_g \in \mathbb{R}^{K}$ and $W_p \in \mathbb{R}^{K}$ are 
parameters to transform the merchant embeddings to gradient (slope) and intercept.
The activation function $\mathrm{SOFTPLUS}(\cdot)$ is used to guarantee the 
monotonic linear mapping between treatment $c$ and objective $f(i,c)$.

Note that the inferred gradient $g_i$ of each merchant $i$ in our model can be 
used to measure the sensitivity to incentive of merchant $i$. A small value of 
$g_i$ indicates that $i$ is less sensitive to the marketing, while a greater
value of $g_i$ indicates the opposite.

To summarize, we have the following objective to optimize:
\begin{align}
\arg\min_{\beta} \sum_{i,c} \ell(f(i,c; \beta), y_{i,c})
\end{align}
where $\ell(\cdot)$ denotes the mean absolute error, and $y_{i,c}$ denotes
the observed commercial objective of merchant $i$ under treatment $c$.
We aim to optimize parameters $\beta = \{W, W_x, W_e, W, \theta, W_g, W_p\}$ in 
an end-to-end model.
We illustrate the architecture of our model in Figure~\ref{fig:arch}.
In practice we use the ADAM~\cite{kingma2014adam} to optimize the above objective 
in a stochastic mini-batch manner,
and the model is implemented with tensorflow~\cite{Abadi2016TensorFlow}.

\subsection{Optimal Incentive as Linear Programming}\label{sec:lp}
As other price optimization problems, in this section, we show how 
we formalize incentive optimization as a linear programming problem
given the estimates of $f(i,c)$.

In the online environment, we need to choose the best treatment for
every merchant. Since the estimates of each merchant's ``gradient'' is positive
due to the $\mathrm{SOFTPLUS}(\cdot)$ activation function, the best strategy
is always choosing the maximum treatment for all the merchants, so that
we can maximize the commercial objective. However, under a limited budget, the 
strategy should consider the value of inferred gradient. Let us denote the 
best strategy $a^*(\cdot)$ for merchant $i$ is treatment $k$, then we have 
$a^{*}(i,c=k) = 1$ and $a^*(i,c=k^-)=0$ for the rest strategies. 
We formalize the following optimization problem:
\begin{align}
\max \sum_{i,c} f(i,c) a^*(i,c), \\
s.t. \sum_{i,c} a^*(i,c) * c \le B, 
\end{align}
then we have the optimal solution $k = \arg\max_{c} f(i,c) - c*\lambda$ 
for each $i$, where $\lambda$ denotes the dual optimal.

We test our strategies in the online environment at Alipay, and show 
the online results in section~\ref{sec:online}.

\section{Experimental Results}
In this section, we conduct extensive empirical experiments to study
the performance of our models. 
We first analyze the offline results based on the dataset
collected in our online experiments.
Next, we show the online results after deploying our approach at Alipay
compared with a multi-layer perceptron neural network model, using the same
optimal strategy as described in section~\ref{sec:lp}.

Note that due to the policy of information sensitivity at Alipay, 
we will not reveal the detailed numbers that could be sensitive.

\subsection{Experimental Settings}
In this section, we will introduce our dataset for learning the model,
and related experimental settings.

\subsubsection{Dataset}
Our two experimental datasets are collected from our online 
experiments (mentioned in section~\ref{sec:samples}) launched at Alipay in two seperated time periods (15 
consecutive days of each), respectively. 
To collect each of the experimental datasets, $0.2\%$ of random 
selected merchants were assigned to
an experimental bucket that places a fixed treatment. We set up a total of $13$ experimental 
buckets for observing merchants with different treatments. That is, each bucket corresponds 
to a distinct treatment in our online experiments, and is responsible for providing the
merchants under control a fixed amount of commissions. Thus, we can assume that the 
merchants with different sensitivities to the marketing campaign are randomly sampled
in different buckets with different treatments.

There are more than 2 millions of labeled 
merchants with observed commercial objects (labels or measures include: the number of payments the merchant
can accomplish in the next 3 days, the number of days the merchant have payments in next 3 days) under given treatments. 
We aim to regress over the commercial 
objectives of each merchant. We build our model on top of the transaction 
network. The edges of the transaction network consist of the transactions 
associated with those labeled merchants in 2-hops. That is, the 
transactions made between the labeled merchants and customers directly, 
and all the transactions made by those customers. This leads to more than 90 
millions of nodes (including merchants and customers) and 
hundred millions of edges (transactions). We summarize two datasets
used for offline experiments in table~\ref{tb:data}.
We collect two different periods of data from online experiments for 
evaluating the robustness of our results.


\subsubsection{Comparison Methods}
To verify the effectiveness of our proposed method, we compare our approach with 
classic regression approaches. It is well known that tree based models~\cite{chen2016xgboost},
linear models, and neural networks~\cite{schmidhuber2015deep} are widely used 
for regression problems. Because the majority 
of features we built are sparse features, and tree based models cannot exploit 
sparse features well, we choose linear regression 
and DNN model (utilizing a multi-layer perceptron architecture)
with the monotonic linear mapping as the output layer
introduced in section~\ref{sec:monotonic}. 
We name our proposed model as GE model in the following experiments.

For DNN, we set the depth of the neural architecture as 2 with latent embedding size
as 256. We set the depth of our graph neural network architecture as 2 (
so that each labeled merchant can aggregate signals from 1-hop neighbors, i.e. their directed customers, 
and 2-hop neighbors, i.e. those merchants who shared the same group of customers) with 
latent embedding size as 256 as well. The rest hyperparameters 
(such as learning rate and regularizers) of the comparison methods 
are tunned by grid search.

We randomly sample 80\% of the merchants for training and the 
remaining for testing respectively from each dataset. 

\subsection{Regression Results}\label{sec:regression}
We conduct experiments on the real-world dataset collected from the online experiments
run by Alipay as described above. To evaluate the performance of the models, 
we choose MAE (Mean Absolute Error) 
and MSE (Mean Square Error) as our metrics, which are commonly used for evaluating 
regression methods. Even though such metrics cannot directly measure how well our approach 
can fit the sensitivities of each merchant, they still reflect how well
the models fit our observed measures of each merchant under specific treatments.

\begin{table}
\begin{center}
\caption{MAE of Linear Regression with our proposed method}\label{tab1}
\begin{tabular}{cC{2cm}C{2cm}}
\toprule
Dataset 1 &  MAE & MSE\\
\midrule
LR &  {0.1432} & {0.0415} \\
DNN &  {0.1404} & {0.0433} \\
GE & {\bfseries{0.1357}} & \bfseries{0.0400} \\
\midrule
Dataset 2 &  MAE & MSE\\
\midrule
LR &  {0.1441} & {0.0417} \\
DNN &  {0.1409} & {0.0441} \\
GE & {\bfseries{0.1361}} & \bfseries{0.0408} \\
\bottomrule
\end{tabular}
\end{center}
\end{table}

As shown in Table~\ref{tab1}, DNN model outperforms Linear Regression in terms of 
MAE, however is doing worse on MSE. This shows that DNN model is more robust due to
the final monotonic linear mapping layer (i.e. successfully reduce the risk
of high variance), but may mislead the characteristics of 
different merchants due to the limited representation learning capacity
(i.e. cannot well captured merchants with subtle differences).
Our GE model produces much better estimates compared with all the other comparison approaches
in terms of both MAE and MSE, which implies that graph representation indeed help 
the representations of merchants and simultaneously depict the merchant's reactions 
to treatments.

\begin{figure}
\centering
\includegraphics[width=0.42\textwidth]{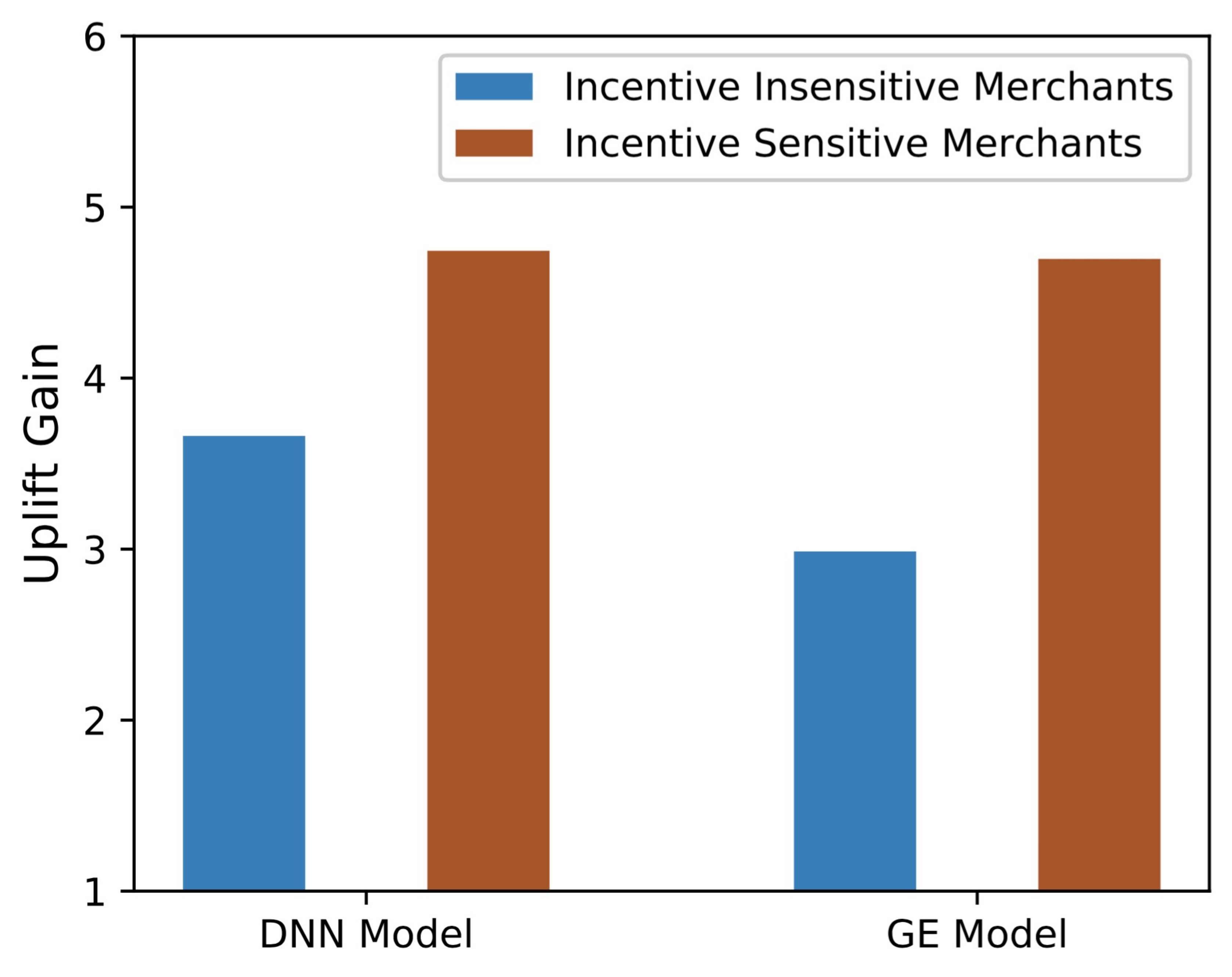}
\caption{Objective-Incentive Curve Analysis} \label{fig3}
\end{figure}

\subsection{Objective-Incentive Curves Analyses}
In this section, we analyze the estimates of so called ``gradient'' defined
in Eq.~\eqref{eq:mapping}. ``Gradient'' can be used to depict how the merchant is sensitive
to the marketing campaign, and describe the objective-incentive curve. 
An accurate estimation of ``gradient'' of each merchant is exactly our goal.

Different from the quantitative results reported in section~\ref{sec:regression},
the estimated ``gradient'' of each merchant is hard to evalute directly.
However, a reasonable assumption is that if a model is well estimated,
the group of merchants with larger estimated ``gradient'' should be able to 
achieve much better results on expectation in terms of commercial objectives, compared
with those merchants with smaller estimated ``gradient'', while all the merchants
are under the same treatment of strong incentives.

We denote $y_h$ as merchants' commercial objective under a treatment with 
a larger amount of incentives, and $y_l$ as that under a treatment with a 
lower amount of incentives. We denote the uplift gain as $u = y_h-y_l$. 
The uplift gain for the merchants who are sensitive to incentives (i.e. with greater
``gradient'') should be comparatively greater than that for less sensitive merchants
(i.e. with smaller ``gradient''). 

To conduct the experiments, we first infer the ``gradient'' for each merchant 
in the test data. By sorting the 
merchants by ``gradient'' in decending order, we can separate the merchants 
into two groups, the incentive sensitive group and incentive insensitive group, 
denoted as $\Omega_+$ and $\Omega_{-}$ respectively. To evaluate an incentive 
optimization model, we need to inspect if the $\Omega_+$ and $\Omega_-$ are 
separated properly.

\begin{figure*}
\centering
\includegraphics[width=0.8\textwidth]{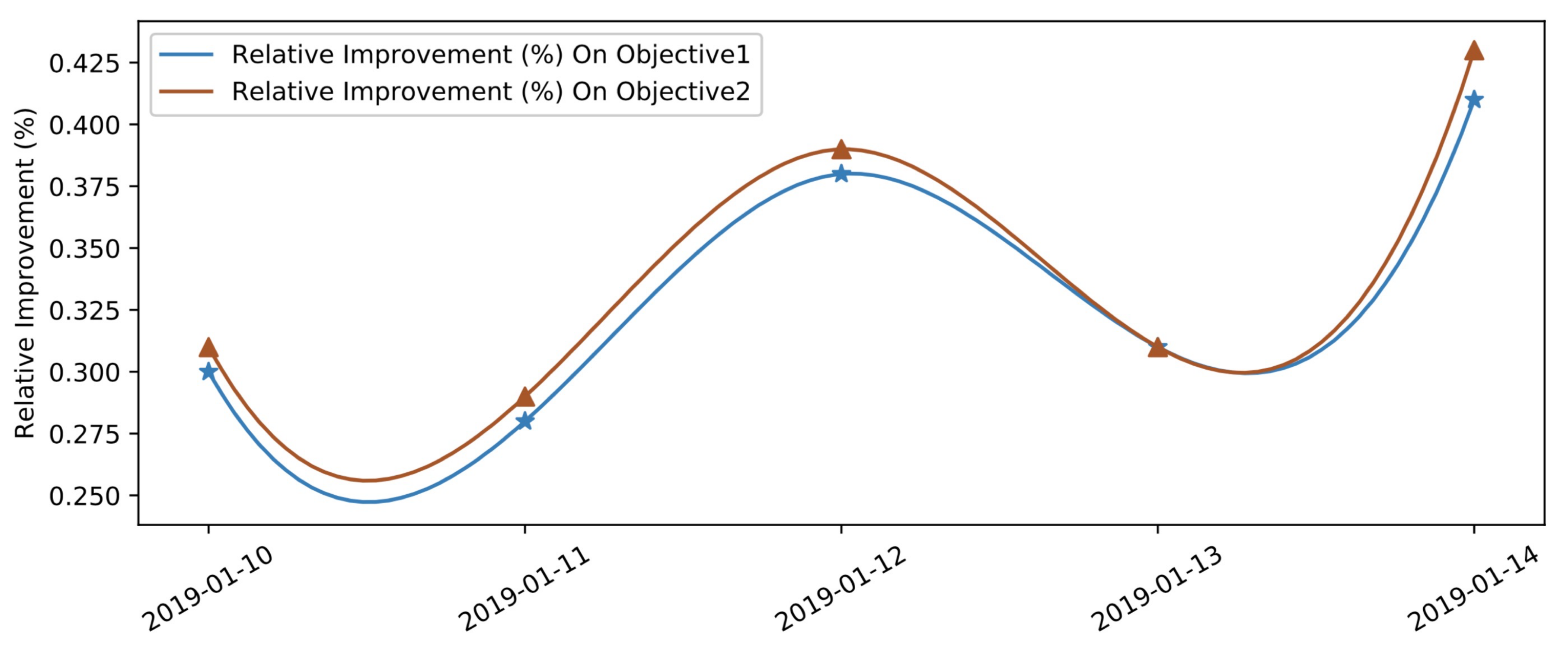}
\caption{Relative improvement on commercial objectives (30\% traffic).} \label{fig4}
\end{figure*} 

For the incentive sensitive group of merchants $\Omega_{+}$, we calculate the 
corresponding uplift gain as $u_+ = y_{h+} - y_{l+}$, where $y_{h+}$ and $y_{l+}$
denote the commercial goals of merchants under treatment of high and low
incentives respectively. Similarly we have $u_-$ as the uplift gain obtained
from insensitve group of merchants. The greater the difference of the uplift gains
$u_+-u_-$ is, the better the model should be. We illustrate the values
of $u_+$ and $u_-$ inferred from both DNN and GE models in Fig.~\ref{fig3}. 
We show the value of $u_+$ using brown bars, and the value of $u_-$ with blue bars.
The gap between two bars of GE model is significantly greater than that of dnn 
model, which implies that the GE model has learned better merchants' 
sensitivity to incentives.


Note that 
the linear regression model cannot produce 
personalized ``gradient'', and we will not report the results accordingly. 

\begin{table}
\begin{center}
\caption{Incentive Sensitivity Comparisons}\label{tab3}
\begin{tabular}{cC{1cm}C{1cm}}
\toprule
Level of Incentive Sensitivity (desc)  & DNN & GE\\
\midrule
$\sim$ 20\% & {5.851} &  {\bfseries{5.935}}\\
20\% $\sim$ 40\% & {2.657} & {\bfseries{4.749}}\\
40\% $\sim$ 60\% & {3.619} & {3.912}\\
60\% $\sim$ 80\% & {3.954} & \bfseries{3.782}\\
80\% $\sim$ & {2.694} & \bfseries{2.157}\\
\bottomrule
\end{tabular}
\end{center}
\end{table}

For DNN and GE model, we sort the merchants in test data by the inferred ``gradient''
in decending order, and separate the merchants into five groups equally. 
We show the uplift gain of each group in Table~\ref{tab3}. 
For the most sensitive merchants, GE model produces a better uplift gain
compared with the DNN model. For the least sensitive 20\% merchants, 
uplift gain produced by GE model is relatively smaller compared with the DNN 
model, which implies that GE model inferred a relative flat objective-incentive
curve for those insensitive merchants.

\newcommand{\tabincell}[2]{\begin{tabular}{@{}#1@{}}#2\end{tabular}}
\begin{table}
\begin{center}
\caption{Relative improvement (\%) of our proposed GE model versus the DNN model (30\% traffic).}\label{tab2}
\begin{tabular}{|c|c|c|c|}
\hline
Models & Cost (\%) & Objective1(\%) & Objective2(\%)\\
\hline
Baseline &  {-} & {-} & {-}\\
GE & {\tabincell{c}{\bfseries{-2.71}\%\\ \lbrack-3.06\%,-2.36\%\rbrack}} & {\tabincell{c}{\bfseries{+0.28}\%\\ \lbrack0.04\%,0.52\%\rbrack}} & {\tabincell{c}{\bfseries{+0.29}\%\\ \lbrack0.04\%,0.54\%\rbrack}}\\
\hline
\end{tabular}
\end{center}
\end{table}

\subsection{Online Results}\label{sec:online}
We deployed our model in the mobile payment marketing scenario at Alipay with a 
standard A/B testing configuration. We conduct our A/B testing experiments beginning 
with a relatively small traffic, i.e. influencing 1\% of merchants, and observe
the measures in consecutive 5 days. After that, we gradually increase the traffic
from 1\%, to 2.5\%, 5\%, 15\%, and observe the measure in 5 days respectively. 
Finally, we conduct the A/B experiments
with traffic 30\% from January 10th to January 14th, and make the final decision on deploying the model
at Alipay. The online results lasting nearly one month consistently show 
that our GE model outperforms the comparison model while we were increasing the traffic.

There exists millions of users who visit our App during the period while we 
were conducting A/B experiments with 30\% traffic. 
In addition to the cost of the marketing campaign, two commercial objectives
(Objective 1: the number of payments averaged over merchants, Objective 2: 
the number of days having at least one payments averaged over merchants) 
are shown as metrics in our experiments. Table~\ref{tab2} shows the 
relative improvements of GE model compared to the DNN model using 30\% of 
traffic. Along with the relative improvements, we also show the confidence intervals
with 95\% confidence level. Compared with the DNN model, our proposed 
approach significantly saved 2.71\% 
cost while improving the two commercial objectives statistical significantly 
(with p-value far less than 0.05).

Figure~\ref{fig4} shows the trends of relative improvements on the two 
commercial goals by applying the strategies optimized using our approach. 
The figure tells that our approach has been gradually improving the overall 
commercial goals given our optimal strategies. This is because each merchant 
is aware of the treatment after they have observed an accumulation 
of operation commissions.

\section{Conclusions}
In this paper, we show our best experiences in the mobile payment marketing 
problem at Alipay. To our best knowledge, this is the first graph neural 
network based marketing experience shared by industry and has been deployed on 
the real-world large scale 
marketing system. To promote the mobile payment activities, we propose
to identify merchants that are sensitive to the marketing campaigns
based on a graph representation learning approach over transaction
networks. We further place a monotonic linear mapping function to
reduce the potential high variance due to limited samples of treatments.
We develop uplift gains as a novel metric to measure the goodness of 
our model, that saves the cost of deploying and observing 
non-effective models online. Our online results in a standard A/B testing configuration
lasting nearly a month shows that our approach is effective.

\end{document}